\documentclass[%
reprint,
superscriptaddress,
 amsmath,amssymb,
 aps,
]{revtex4-1}

\usepackage{graphicx}
\usepackage{dcolumn}
\usepackage{bm}
\usepackage{soul} 

\usepackage{mathtools} 

\usepackage{listings} 
\usepackage[flushleft]{threeparttable} 

\usepackage{dsfont}
\usepackage{color,psfrag}

\usepackage{xcolor} 

\def\mathcolor#1#{\@mathcolor{#1}}
\def\@mathcolor#1#2#3{%
  \protect\leavevmode
  \begingroup
    \color#1{#2}#3%
  \endgroup
}

\newcommand{\mat}[1]{\mathbf{#1}} 

\newcommand{\tran}{^{\mathstrut\scriptscriptstyle\top}} 

\begin{document}

\preprint{APS/123-QED}

\title{sGDML: Constructing Accurate and Data Efficient Molecular Force Fields Using Machine Learning}

\author{Stefan Chmiela}
 \affiliation{Machine Learning Group, Technische Universit\"at Berlin, 10587 Berlin, Germany}

\author{Huziel E. Sauceda}
 \affiliation{Fritz-Haber-Institut der Max-Planck-Gesellschaft, 14195 Berlin, Germany}

\author{Igor Poltavsky}
\affiliation{Physics and Materials Science Research Unit, University of Luxembourg, L-1511 Luxembourg, Luxembourg}

\author{Klaus-Robert M\"uller}%
\email{klaus-robert.mueller@tu-berlin.de}
\affiliation{Machine Learning Group, Technische Universit\"at Berlin, 10587 Berlin, Germany}
\affiliation{Department of Brain and Cognitive Engineering, Korea University, Anam-dong, Seongbuk-gu, Seoul 02841, Korea}
\affiliation{Max Planck Institute for Informatics, Stuhlsatzenhausweg, 66123 Saarbr\"ucken, Germany}

\author{Alexandre Tkatchenko}
 \email{alexandre.tkatchenko@uni.lu}
\affiliation{Physics and Materials Science Research Unit, University of Luxembourg, L-1511 Luxembourg, Luxembourg}

\date{\today}

\begin{abstract}

We present an optimized implementation of the recently proposed symmetric gradient domain machine learning (sGDML) model. The sGDML model is able to faithfully reproduce global potential energy surfaces (PES) for molecules with a few dozen atoms from a limited number of user-provided reference molecular conformations and the associated atomic forces. 
Here, we introduce a Python software package to reconstruct and evaluate custom sGDML force fields (FFs), without requiring in-depth knowledge about the details of the model. A user-friendly command-line interface offers assistance through the complete process of model creation, in an effort to make this novel machine learning approach accessible to broad practitioners. Our paper serves as a documentation, but also includes a practical application example of how to reconstruct and use a PBE0+MBD FF for paracetamol. Finally, we show how to interface sGDML with the FF simulation engines ASE (Larsen et al., J. Phys. Condens. Matter 29, 273002 (2017)) and i-PI (Kapil et al., Comput. Phys. Commun. 236, 214-223 (2019)) to run numerical experiments, including structure optimization, classical and path integral molecular dynamics and nudged elastic band calculations.

\end{abstract}

\maketitle

\section{Introduction}

Machine learning has had a transformative effect on the modeling of highly accurate potential energy surfaces (PES), offering \emph{ab initio} accuracy at the computational cost in between that of classical interatomic potentials and density-functional approximations to the exact solution of the Schr\"odinger equation. This development is propelled by numerous significant improvements in molecular representation~\cite{Rupp2012,Hansen2013,Hansen2015,Rupp2015,
eickenberg2018solid,Ceriotti2016,artrith2017efficient,Ceriotti2017,Glielmo2017,yao2017many,john2017many,faber2017prediction,glielmo2018efficient,tang2018atomistic,Grisafi2018,pronobis2018many}, inference approaches~\cite{Bartok2010,Bartok2013,Behler2007,Behler2012,Montavon2013a,Bartok2015_GAP,Ramprasad2015,Tristan2015,Behler2016,Brockherde2017,Gastegger2017,Schutt2017,huang1707dna,huan2017universal,schutt2017schnet,Tristan,zhang2018deep,lubbers2018hierarchical,ryczko2018convolutional,kanamori2018exploring,hy2018predicting,Smith2018,Clementi2018,winter2019}, data sampling schemes~\cite{DeVita2015,Shapeev2017,dral2017structure,noe2018,noe2018b}, as well as new explanation methods~\cite{innvestigate,meila2018} and software implementations~\cite{yao2018tensormol,schnetpack2018} that make these advances practical and widely available.

In this work we present an optimized implementation of the recently proposed sGDML model~\cite{gdml, gdml2, Sauceda2019}, which is able to achieve high data efficiency through the incorporation of spatial and temporal physical symmetries of molecular systems into a gradient-domain machine learning approach. Unlike traditional FFs, this global model imposes no hypothesized interaction pattern on the nuclei and is thus suited for describing any complex physical interaction. 
The sGDML model can reach spectroscopic accuracy in the energy for small molecules like benzene and toluene, and an accuracy of a few wavenumbers for the position of the spectral peaks. It calculates energies and forces at speeds around four and eight orders of magnitude faster than DFT and CCSD(T), respectively. Compared to conventional FFs, sGDML is only one to three orders of magnitude slower. This brings it is closer to polarizable force fields~\cite{benoit2011} than classical force fields like AMBER~\cite{amber18}, CHARMM~\cite{charmm2009}, or GROMACS~\cite{gromacs2005} in terms of speed.

We have demonstrated previously that sGDML enables converged MD simulations at quantum-chemical CCSD(T) level of accuracy for flexible molecules with up to a few dozen atoms~\cite{gdml2}. Such simulations are key for accurate predictions of molecular behavior at realistic conditions, but unfeasible within brute-force \emph{ab initio} approaches since they would require millions of CPU years.

Here, we describe a Python package that provides user-friendly routines to reconstruct and query sGDML models, based on a small set of reference geometries with corresponding forces and energies as the only input. Forces and energies for new geometries can then be queried in a fraction of a millisecond on a regular laptop computer (see Table~\ref{tab:performance}). From the user point of view, sGDML models behave like traditional FFs, with the added benefit of reproducing the accuracy of the provided reference data. A variety of highly redundant PES sampling and exploration tasks can now be performed, including molecular dynamics, vibrational analysis, structure optimization and the computation of transition paths.

We demonstrate the use of our software on the example of reconstructing the PES for the paracetamol molecule, which we then interface with the popular i-PI code~\cite{ipiv2} to perform a MD simulation of this molecule.

\begin{figure}[ht]
    \centering
    \includegraphics[width=1.0\columnwidth]{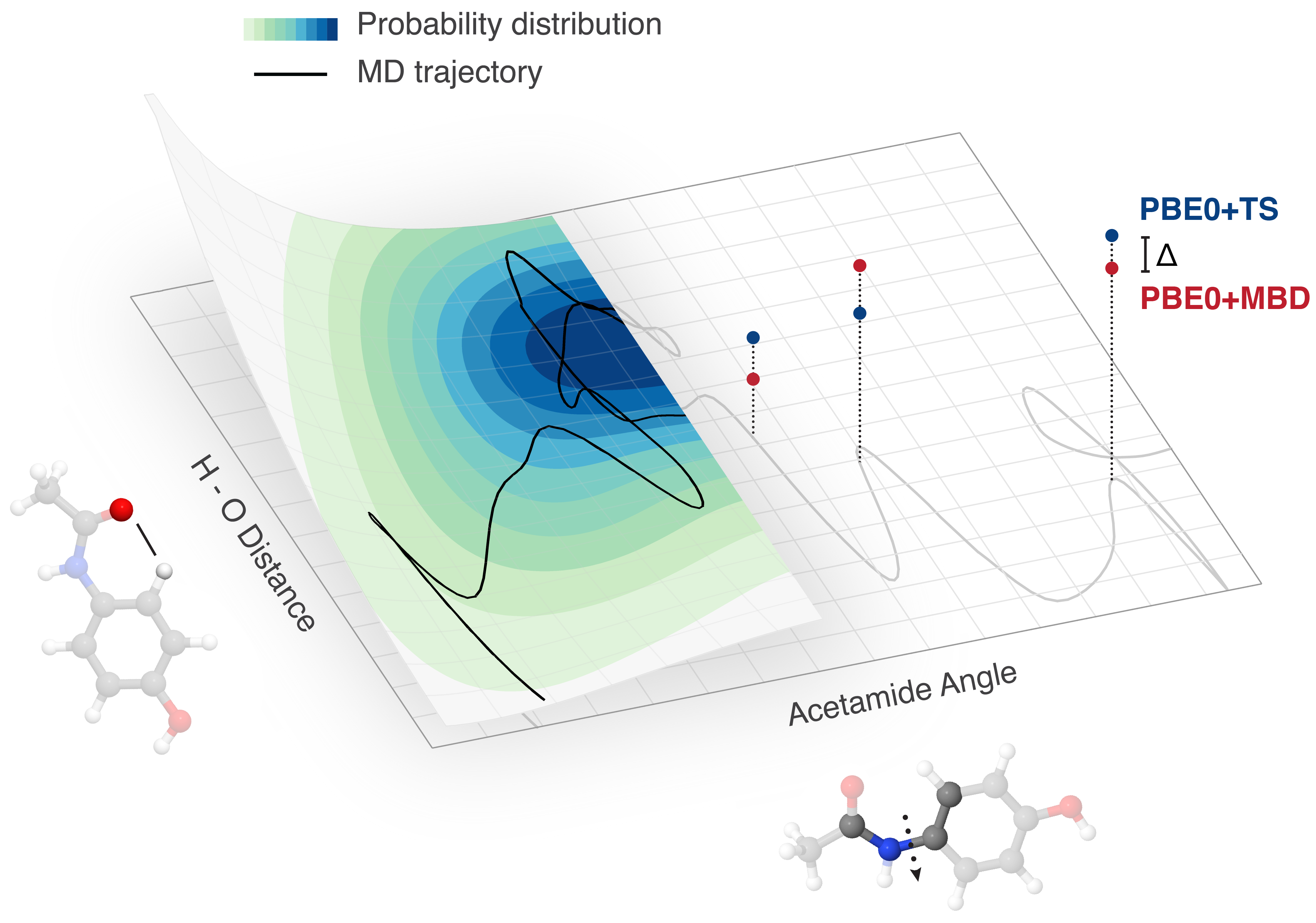}
     \caption{Reference data generation (paracetamol): Geometries are sampled from a sufficiently long, but cheap DFT-PBE+TS MD trajectory to ensure optimal coverage of the configuration space. Energy and force labels for this small subset of the trajectory are then recomputed at the higher DFT-PBE0+MBD level of theory and used for training the sGDML model. The full PES will be reconstructed at the accuracy of the DFT-PBE0+MBD reference data.}
    \label{fig:sampling}
\end{figure}

\section{Program Overview}\label{overview}

Our main goal with this reference implementation of sGDML is to provide a compact working example of the model in an accessible programming language. We offer one variant of our program with sophisticated parallel processing support for ubiquitous multi-core CPUs and another one for state-of-the-art multi-GPU computing environments. While adhering to best-practices for writing readable code, our main focus is on performance. Hence, we make full use of programming language specific optimizations, e.g. vectorized operations as a replacement for slow nested loops. These allow us to achieve performance comparable to natively compiled code.

The tasks of FF reconstruction and evaluation are separated into independent modules for training (\texttt{train}) and prediction (\texttt{predict}). All necessary routines for reference data sampling, symmetry recovery, and model parametrization are packaged in the training module. It generates lightweight model files that contain the preprocessed essentials for FF evaluation, which are then independently instantiated and queried using the second module. Such separation makes it possible to centralize training on a high performance computer while the completed model can be efficiently used anywhere. For that purpose, we designed the prediction module to be minimal and self-contained in the sense that it only contains logic that is absolutely essential for generating energy and forces for a given input geometry. This greatly simplifies the integration of sGDML into any application that requires a FF.

On top of that, we include a user-friendly command-line interface (CLI) \texttt{sgdml} that exposes the functionality of both modules to the shell. It provides an easy introduction to sGDML model reconstruction, guiding the the user through the complete process. To get started, it is not required to be familiar with the intricacies of the theory behind sGDML, which is why our software also provides a good entry point for newcomers to the field.

\subsection{User Input}\label{UI}

The essential ingredient for training and validating a sGDML model is a user-provided reference dataset, specifically a set of Cartesian geometries with corresponding total energy and atomic-force labels. Those labels can be generated from any level of theory, e.g. \textit{ab-initio} calculations, any method derived from DFT (e.g. Kohn-Sham or other orbital-free variants) or even regular FFs, since the sGDML model is not biased towards a specific kind of reference data. Force labels are needed, because our approach implements energy conservation as an explicit linear operator constraint, by modeling the FF reconstruction $\mathbf{\hat{f}_F} = -\nabla\hat{f}_E$ as the transformation of an underlying energy model~\cite{gdml}. Force learning affords data-efficiency advantages, as they are more informative per example, while being generally cheaper to compute analytically than collecting the same derivative information via numerical approximation from energy examples. Since forces are true quantum-mechanical observables, they preserve all information regarding the quantum nature of the system and therefore pass it on to the model.

A key consideration when composing a reference dataset, is the choice of sample region on the PES. Generally, we want to keep the covered area tight, avoiding the inclusion of configuration space that will not be explored in the specific application of the trained model. With that being said, we also aim to limit the need for extrapolation, which usually carries a performance penalty. All isomeric conformers of interest, including the transition pathways, need to be well represented in the dataset.

The sGDML model is unit-agnostic, meaning that the energy and force predictions will simply inherit the units of the training labels. Particular attention should be paid to ensuring that the unit of force (e.g. kcal $\text{mol}^{-1} \text{\r{A}}^{-1}$) is consistent with the unit of energy (e.g. kcal $\text{mol}^{-1}$) and the unit of length (e.g. \r{A}) used in the provided energy labels and geometries, respectively.
While the model will quietly convert different length units between input and output, it is not able to adapt the energy unit. As a good practice, we strongly advise against mixing units in the same dataset, since an implicit unit conversion within the trained model is not a behavior that the user expects.

All geometries within a dataset must use a consistent atom indexing and every derived model should be queried using the same order. This is because the invariance of sGDML models is restricted to permutational symmetries that are physically feasible and statistically relevant, which does not include the full symmetry group of the molecule in general. Arbitrarily indexed query geometries may not fall within the set of interchangeable representations and hence yield undefined outputs. While it would be technically straightforward to extend the sGDML prediction routine to support randomly index inputs, we deliberately omitted that functionality in favor of evaluation speed.

We use NumPy binary files as the native file format for our application, but include converters from and to various popular plaintext formats. Support for additional file types can be easily extended, by using one of the included conversion scripts as a template. One of the main reasons for using a custom file format is the inclusion of metadata that makes the origin of each model traceable and data integrity verifiable.

\section{Method}\label{method}

GDML~\cite{gdml} constructs conservative FFs by solving the normal equation of the ridge estimator in the gradient domain, using the Hessian matrix of a kernel function as the covariance structure. This constitutes an explicit gradient operator constraint, which dictates that the reconstructed FF must be a transformation of some unknown underlying energy model (see Supplementary Information).

During training, all partial forces of a molecule are mapped simultaneously
\begin{equation}
\left(\mathbf{K}_{\text{Hess}(\kappa)} + \lambda \mathbb{I}\right) \vec{\alpha}= \nabla V_{BO} = -\mathbf{F} \text{.}
\label{eq:normal_equation}
\end{equation}
The underlying kernel function $\kappa$ is chosen from the parametric Mat\'{e}rn family~\cite{matern1986spatial, Gradshteyn2007, Gneiting2010} (see Supplementary Information). It can be regarded as a generalization of the universal squared exponential kernel with an additional smoothness parameter $n$. Our parameterization $n=2$ resembles the exponential kernel, while being sufficiently differentiable. Cartesian geometries that have undergone translation or rotation are disambiguated by representing them as the matrix of inverse pairwise atom distances with entries $\vec{x} = (\mat{D})_{ij} = \|r_i-r_j\|^{-1}$.

The resulting model is guaranteed to be integrable and the global potential energy hyper-surface (PES) is hence easily recovered as the integral
\begin{equation} 
\int \mathbf{\hat{f}_F} \, \mathrm{d}R = -\hat{f}_E + c \text{,}
\end{equation}
which is defined up to an additive constant $c$. We determine this value in the least-squares sense (see Supplementary Information) using the corresponding energy labels $E$ for each geometry in the training set, even though these are not explicitly included in the objective function.

Building on GDML, we recently proposed the sGDML model~\cite{gdml2}, which additionally incorporates all relevant rigid space group symmetries (e.g. reflection operation), as well as dynamic non-rigid symmetries (e.g. methyl group rotations). Typically, the identification of symmetries requires chemical and physical intuition about the system at hand, which is impractical in a ML setting. Through a data-driven multi-partite matching approach, we automate the discovery of permutation matrices $\mat{P}$ that realize the assignment between adjacency matrices $(\mat{A})_{ij} = \|\vec{r}_i - \vec{r}_j\|$ of molecular graph pairs $G$ and $H$ in different permutational configurations $\tau$,
\begin{equation}
\operatorname*{arg\,min}_{\tau} \mathcal{L}(\tau) = \|\mat{P}(\tau)\mat{A}_G\mat{P}(\tau)\tran - \mat{A}_H\|^2
\label{eq:matching_objective}
\end{equation}
and thus between symmetric transformations undergone within the scope of a dataset~\cite{Umeyama1988}. The resulting approximate local pairwise matchings are subsequently globally synchronized using transitivity as the consistency criterion~\cite{Pachauri2013}. We limit this search to the particular training set of a model and require no additional data. In doing so, we simultaneously exclude combinatorially feasible, but physically irrelevant permutational configurations that are inaccessible without crossing impassable energy barriers. Together, these two types of physical constraints greatly reduce the intrinsic complexity of the FF learning problem.

A machine learning model is only useful, if it is able to generalize to unseen data, once trained~\cite{Bishop2006,Scholkopfa,Muller2001}. Good prediction performance on the training sample is however not indicative of a good performance on new data from the same distribution, as the learning algorithm may have erroneously responded to noise in the data. Typically, this is prevented by penalizing the complexity of the solution via the inclusion of a regularization term in the objective function of the model. As part of the training process, the influence $\lambda \ge 0$ of the regularizer (see Eq.~\ref{eq:normal_equation}) is varied to find the combination with the lowest prediction error on a held-out validation set. In the particular case of sGDML, this so-called \emph{model selection} procedure includes a second parameter for the length scale $\sigma$ of the Mat\'{e}rn covariance function. Traditionally, this highly non-convex optimization is implemented as an exhaustive search on a predefined grid, involving the training of multiple model candidates with varying hyper-parameters choices. The generalization performance of the winning model is then estimated on a third test dataset that is completely independent from the first two ones that participated in the training process.

\section{Usage}

Our program includes a set of convenience routines that assist the user in reconstructing sGDML models from beginning to end. It will walk the user through the complete process of data sampling, symmetry recovery, training with hyper-parameter optimization and testing to generate a ready-to-use model. Greater control over this procedure may be taken by running the involved subroutines individually, either via the CLI or using the Python interface of the \texttt{train} and \texttt{predict} modules (see Supplementary Information). From the CLI, the assisted training process is initiated by simply calling
\begin{lstlisting}[frame=none]
$ sgdml all <dataset_file> \
	      <n_train> <n_validate> [<n_test>] \
	      [--sig <list_or_range>]
\end{lstlisting}
with a path to the reference dataset as the argument. The parameter \texttt{n\_train} specifies how many data points are used for training: larger training sets yield more accurate models, but at increased computational cost (see Supplementary Information). During model selection, the performance of a model candidate is assessed based on the comparison of \texttt{n\_validate} predicted forces and energies with the true labels. Optionally, the number of test points \texttt{n\_test} can be specified, otherwise this parameter will be set to the maximum value for the best possible final estimate of the generalization error. Large validation and test datasets are desirable as they only increase computational cost marginally, while yielding better error estimates. Additionally, the search grid for the hyper-parameter $\sigma$ can be specified as a space-delimited list (\texttt{--sig <s1> <s2> ... <sN>}), or a range of evenly spaced values within a given interval (\texttt{--sig <start>:<step>:<stop>}), or a combination of both.

Training, validation and test subsets are sampled from the provided bulk dataset without overlap, unless individual datasets (\texttt{-v <validation\_dataset>} and/or \texttt{-t <test\_dataset>}) are specified. For optimal prediction performance, it is crucial for the training set to represent the distribution the model will encounter. Likewise, we can only reliably assess its expected generalization error if we validate and test on representative datasets. With the assumption that the bulk dataset adequately describes the molecular configuration space that will be visited in the application of trained model, our sampling method automatically extracts stratified subsets that properly follow the estimated probability energy density function of the full dataset.

\begin{table}

  \centering
  \caption{Training times for various sGDML models based on 1000 reference data using an analytic solver on a Intel Xeon E5-2640 CPU at 2.40GHz. For the same models we also list the force and energy prediction performance for sequential geometry evaluations on a 2.8 GHz Intel Core i7 notebook.}
  \label{tab:performance}
  \setlength\extrarowheight{3pt}
  \begin{ruledtabular}
  \begin{tabular}{lrr}
    Molecule  & \multicolumn{1}{l}{Training [min]}& \multicolumn{1}{l}{Prediction [geom./sec]}\\[0.9ex]
    \hline
    Benzene  & 1.9 & 434.7\\
    Uracil  & 2.0& 1103.9\\
    Naphthalene & 5.8& 446.9\\
    Aspirin  & 9.5 & 295.0\\
    Salicylic acid& 4.7& 894.2\\
    Malonaldehyde  & 2.5& 1001.0\\
    Ethanol  & 2.4& 826.2\\
    Toluene  & 3.6& 326.3\\
    Paracetamol & 7.9 & 208.5\\
    Azobenzene & 17.8 & 182.6\\
  \end{tabular}
  \end{ruledtabular}
\end{table}

\subsection{Training}

Every sGDML model emerges from a \emph{training task}, which is a file that packages the configuration for a particular training run, including the indices of the training and validation data points, the permutational symmetries of the molecule, as well as a particular hyper-parameters choice. A batch of training tasks for a range of hyper-parameters is generated with the \texttt{create}-command
\begin{lstlisting}[frame=none]
$ sgdml create <dataset_file> <n_train> <n_valid> \
                 [-sig <list_or_range>]
\end{lstlisting}
which sets up a directory containing the corresponding task files. All parameters used here have been introduced previously. This routine will sample training and validation datasets form the provided bulk dataset, recover the symmetries in the geometry and package everything into individual tasks for each $\sigma$ in the provided range.

Using the \texttt{train}-command and the task directory created in the previous step, the training process is invoked with
\begin{lstlisting}[frame=none]
$ sgdml train <task_dir_or_file>
\end{lstlisting}
For each training task, this resource intensive process creates a model candidate in the same directory. Alternatively, a path to a single file can be passed to execute an individual task, which is useful when submitting batch jobs to distributed computing environments. Parallelization is easy, because the full training dataset is stored in each task file, so that each training job can be performed in isolation, without referencing the potentially large common bulk dataset. All model candidates are stored in the task directory. In the next step, we will evaluate the performance of each model on the validation set and select the leading hyper-parameter choice.

The validation process is invoked via
\begin{lstlisting}[frame=none]
$ sgdml validate <model_dir_or_file> <dataset_file>
\end{lstlisting}
for the whole directory or individual models. As the validation dataset has been predetermined during training task creation and stored in the model, we must pass the originally referenced dataset, otherwise the program can not continue.

Finally, we keep the best performing model from the full set of candidates based on the lowest root-mean-square error (RMSE), which is the metric used in the objective function for the parameterization of the model.
\begin{lstlisting}[frame=none]
$ sgdml select <model_dir>
\end{lstlisting}
Because the validation dataset was used to determine the optimal hyper-parameters, it participated in the training process, very much like the actual training data. To estimate the generalization behavior of the final model in an unbiased way, we will hence use a third independent test dataset and measure its performance once again by calling
\begin{lstlisting}[frame=none]
$ sgdml test <model_file_or_dir> <dataset_file> \
               [<n_test>]
\end{lstlisting}
The reliability of this estimate can be improved by using as many data points as available. Omitting the last parameter selects all points for the dataset that were not involved in the training process of the model.

\subsection{Prediction}
\label{ssec:gdml-predict}

The sGDML force estimator trained on $M$ reference geometries, each with $3N$ partial derivatives and $S$ symmetry transformations, takes the form
\begin{equation}
\mat{\hat{f}_F}(\vec{x}) = \sum^M_{i} \sum^{3N}_{l} \sum^{S}_{q} (\mat{P}_q \vec{\alpha}_{i})_l  \frac{\partial}{\partial x_{l}} \nabla \kappa(\vec{x},\mat{P}_{q}\vec{x}_i) \text{.}
\label{eq:force_model}
\end{equation}
Due to linearity of integration, the corresponding energy predictor is identical up to the second derivative operator on the kernel function, which allows the simultaneous computation of both quantities without computational overhead. It is easy to see that this expression offers a lot of potential for parallelization, which we fully exploit in our code. The amount of concurrent work performed by our implementation is governed by two optional parameters that depend on the host hardware: the number of parallel processes \texttt{num\_processes} and the chunk size \texttt{chunk\_size} in which data items are processed at once. A chunk refers to a vectorized operation that is passed as one big task to Python's underlying high-performance libraries. Both parameters can be automatically tuned for optimal performance by simply calling 
\begin{lstlisting}[frame=none]
gdml_predict.set_opt_parallelism()
\end{lstlisting}
after instantiation of the prediction class. This routine runs a small benchmark that tests feasible configurations by repeatedly calling the \texttt{predict}-function while measuring execution time. Because this routine takes a few seconds to complete, its runtime is only amortized when followed by a large amount of FF evaluations.

Once a sGDML model is trained, it can be integrated into external programs via the \texttt{gdml\_predict} module. A new model instance is created using
\begin{lstlisting}[frame=none]
gdml_predict = GDMLPredict(model,\
                             [chunk size],\
                             [num_processes])
\end{lstlisting}
Force and energy predictions for a geometry are then simply generated using
\begin{lstlisting}[frame=none]
r,_ = io.read_xyz(geometry_path)
e,f = gdml_predict.predict(r)
\end{lstlisting}
This function also accepts a batch of geometries at once, which is useful in applications where multiple independent geometries need to be computed at the same time, e.g. path integral molecular dynamics with a variety of thermostats and statistical ensembles, or in transition path search.

\begin{figure*}[!ht]
    \centering
    \includegraphics[width=1.0\textwidth]{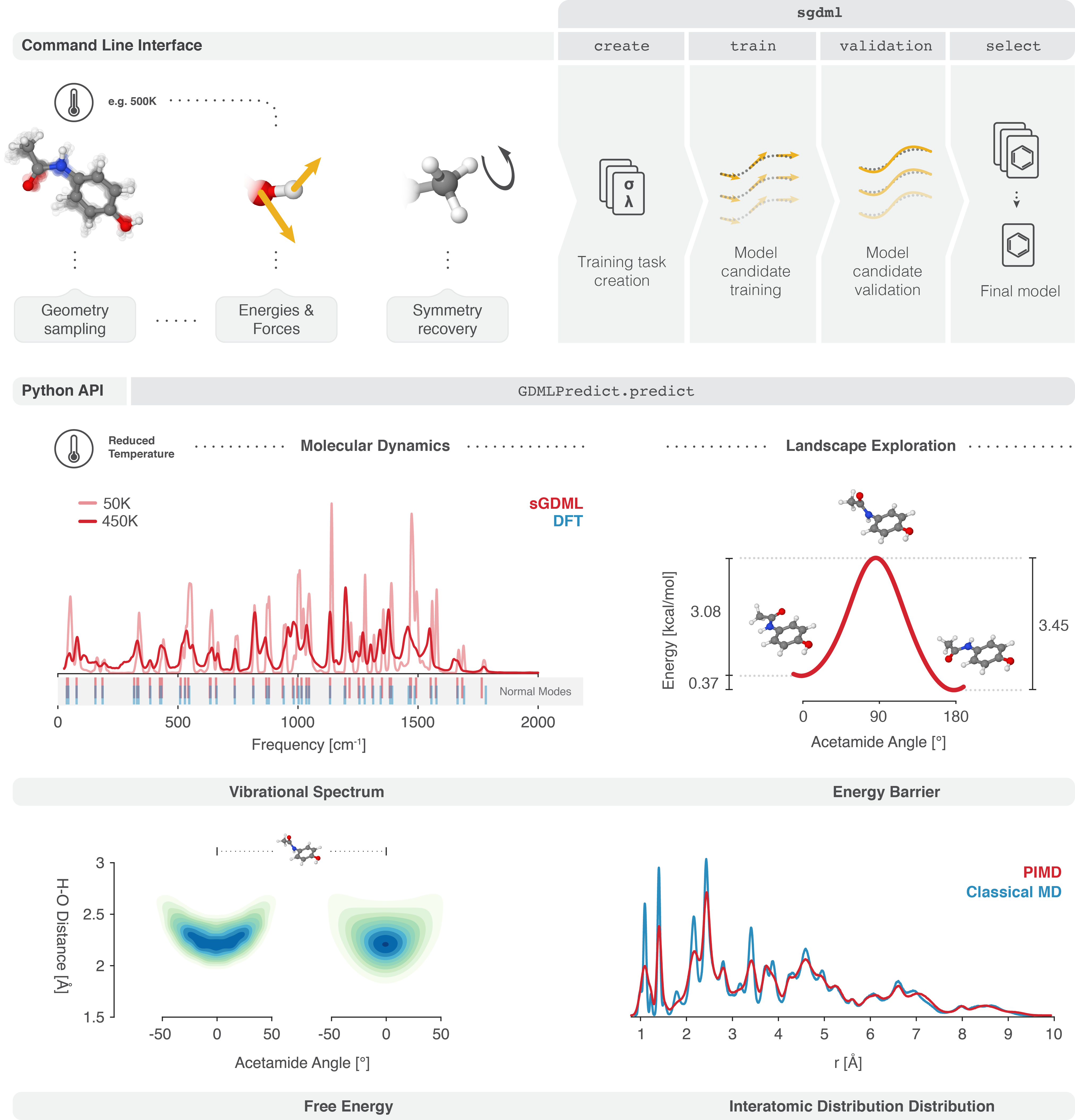}
      \caption{Top: From a provided dataset of molecular geometries with corresponding energy and force labels, our sGMDL implementation creates a fully cross-validated FF model. Bottom: This lightweight model can then be used to speed up various PES sampling intensive applications, like molecular dynamics or the computation of transition paths. Interfacing ASE allows for easy computation of normal modes, vibrational spectra or nudged elastic band optimizations (middle row). Our interface to i-PI enables path integral molecular dynamics simulations (PIMD), which we use to compute the free energies and interatomic distance distributions $h(\vec{r})$ with classical MD and PIMD (bottom row).}
    \label{fig:overview}
\end{figure*}

\section{Example Application: Paracetamol}

To outline the process of FF construction from beginning to end, we consider the paracetamol molecule as an example. Our aim is to create a model for use in long time-scale MD simulations at room temperature (300 K) and an accuracy level of PBE0+MBD. This application is interesting, because a direct sampling at this level of theory would be prohibitively expensive and require hundreds of millions of CPU hours. 

First, we will generate a minimal training set that captures all relevant geometrical configurations. Unreliable predictions are prevented by ensuring that the planned simulations never wander off the regime of configuration space that is covered by training data. In the same vein, we want to exclude sections of the PES that will never be queried in the actual application of the trained model as this would unnecessarily complicate the reconstruction task. 
Here, we use a sufficiently long MD trajectory at a higher temperature of 500 K (see Fig.~\ref{fig:sampling}) to provide the appropriate coverage. The actual training set is then constructed as a small subset of the original trajectory whose energies follow the Maxwell-Boltzmann distribution (see Fig.~\ref{fig:overview}). Foregoing a prohibitively expensive long timescale MD simulation at the theory level DFT-PBE0+MBD with a large basis set, we use a cheap DFT-PBE+TS trajectory as the geometry sampling method and only recompute the corresponding energy and force labels for the small subset of selected training points at the higher level of theory (see Fig.~\ref{fig:sampling}). 

We remark that this sampling scheme is based on the assumption that the PBE+TS energy surface is a good proxy for the topographical structure of the PBE0+MBD surface, as overly strong approximations may yield a sampling profile that misses important features. It is furthermore important to choose a fine-enough time step for the MD simulation, so that the relevant areas of configuration space are sampled with correct probability. As a rule of thumb we use one tenth of the period of the highest frequency oscillator in the system (i.e. hydrogen stretching frequencies). For example, if the highest vibration frequency in paracetamol is 3600 wavenumbers (i.e. period of $9.3$ fs), then our time step works out to $\sim{1}$ fs.
We have obtained the simulated trajectory as a dataset file in \emph{extended} XYZ format, which contains our collected geometries with corresponding forces in additional columns and the energy labels in the comment line. The next step is to convert it to the native sGDML binary format, which is the basis for all forthcoming steps:
\begin{lstlisting}[frame=none]
$ sgdml_dataset_from_xyz.py paracetamol.xyz
\end{lstlisting}
With the resulting dataset file \texttt{d\_paracetamol.npz}, we will now run the fully automated sGDML training assistant which will walk us through all steps necessary to obtain a fully trained and tested model:
\begin{lstlisting}[frame=none]
$ sgdml all d_paracetamol.npz 1000 500
\end{lstlisting}
We have chosen to reconstruct the PES using $1000$ training points, sampled from the provided dataset file, and to use $500$ separate geometries to validate the performance of our candidates during model selection. We omit the argument for the number of test data points, as we want the program to test the resulting model on all remaining data points from the set. The assistant will now automatically split the dataset, train models for a series of hyper-parameter candidates, validate all models, select the most accurate one, finally test it and output a model file \texttt{m\_paracetamol.npz}. Using only this file, we can easily use the newly reconstructed paracetamol force field in existing applications:
\begin{minipage}[c]{0.95\columnwidth}
\begin{lstlisting}[frame=none]
import numpy as np
from sgdml.predict import GDMLPredict

model = np.load('m_paracetamol.npz')
gdml = GDMLPredict(model)
\end{lstlisting}
\end{minipage}
and make predictions using
\begin{lstlisting}[frame=none]
e,f = gdml.predict(r)
\end{lstlisting}
Interfaces to two popular FF simulation engines are already included with our software package: a \texttt{Calculator} for ASE~\cite{ase17} and a i-PI~\cite{ipiv2} \texttt{ForceField}-object. ASE enables various standard simulation tasks including structure optimization, vibrational analysis, molecular dynamics simulations and nudged elastic band calculations, whereas i-PI implements path integral MD to study molecular phenomena that are driven by nuclear quantum effects and a wide variety of sophisticated methods to compute quantum observables~\cite{ipiv2} (see Supplementary Information). In the following, we present in step-by-step fashion how to integrate sGDML with ASE and i-PI and demonstrate practical applications for which it is useful.
 
\subsubsection{ASE: Normal mode analysis}

We will now proceed with a normal mode analysis of paracetamol using ASE. After attaching the \texttt{SGDMLCalculator} to the \texttt{Atoms}-object, we relax an initial geometry \texttt{paracetamol.xyz} with the BFGS optimizer. Then we simply calculate the vibrational modes in the harmonic approximation using \texttt{Vibrations}:
\begin{lstlisting}[frame=none]
from sgdml.intf.ase import SGDMLCalculator

from ase.io.xyz import read_xyz
from ase.optimize import BFGS
from ase.vibrations import Vibrations

mol = read_xyz('paracetamol.xyz').next()

sgdml = SGDMLCalculator('m_paracetamol.npz')
mol.set_calculator(sgdml)

vib = Vibrations(mol)
vib.run()
vib.summary()
vib.write_jmol()
vib.clean()
\end{lstlisting}

This process will output a table with all vibrational frequencies, but also write a file \texttt{vib.xyz} that can be imported into e.g. Jmol to visualize the vibrational modes. To validate the accuracy of our normal mode frequencies, we compare directly with the spectrum from DFT-PBE0+MBD using FHI-aims. Fig.~\ref{fig:overview} outlines the difference between the two sets of normal mode frequencies showing a maximum deviation of only $\sim$4 cm$^{-1}$. This result evinces the robustness of our model given that no explicit information was provided regarding the normal modes.

\subsubsection{i-PI: Molecular dynamics}
In physics and chemistry many of the molecular phenomena, are driven by nuclear quantum effects (NQE), in particular for protons, this nuclear delocalization gives rise to numerous quantum phenomena e.g. zero-point energy and tunneling. Different methods have been developed to incorporate such effects in the BO approximation, path integral molecular dynamics (PIMD) being one of the most widely used. The i-PI software offers an efficient PIMD implementation including state-of-the-art integrators and thermostats~\cite{ipiv2}. The sGDML model can be easily incorporated in i-PI as a force and energy provider \texttt{class FFsGDML()} (see Supplementary Information for details on the interface).
Once the sGDML force field is available in i-PI, running a MD simulation is straightforward. A minimal set up requires the initial coordinates \texttt{paracetamol.xyz}, the sGDML model file \texttt{m\_paracetamol.npz} and the input file \texttt{input.xml} which specifies the parameters of the simulation, e.g. force field, ensemble, temperature, thermostat, integration step, etc. Then running the MD simulations requires just one simple command: \texttt{python i-pi input.xml}.

From these MD simulations, we can compute a wide variety of properties such as finite temperature vibrational spectra, free energy surfaces, radial distribution functions, energies, heat capacities, etc. As an example, we analyze the effect of the temperature on the vibrational spectrum. Fig.~\ref{fig:overview} shows the comparison of the normal modes and vibrational spectra at different temperatures (50K and 450K) using classical MD simulations. From this comparison, the effect of the anharmonicities at high temperatures is evident, given the noticeable red-shift in the frequency peaks. Beyond classical MD, we can explore the NQE by running PIMD in i-PI. An important measure of the NQE is the interatomic distance distributions, $h(r)$, shown in Fig.~\ref{fig:overview}. The deviation between the two curves for classical MD and PIMD gives the magnitude of the delocalization of mean pair distances. This analysis provides an idea of the delocalization of the atomic nuclei in the molecule due to NQE.

\section{Conclusion}

We developed and described a Python based reference implementation of sGDML to provide a versatile tool for highly accurate and data efficient machine learning-based FF estimation. On top of the core sGDML libraries, we provide a user-friendly CLI to assist with data preparation and model creation. The complete reconstruction process leading up to model integration into a MD simulation environment is demonstrated on a practical example that serves as a blue print for practitioners. 

The sGDML approach is able to exceed the capabilities of traditional FFs in the study of small to medium-sized molecules significantly, as its flexible functional form gives it the expressiveness to model complex covalent and non-covalent interactions and thus to capture the thermodynamical properties of the system.
Those interactions are essentially modeled at the accuracy of the reference data that is provided by the user. Our implementation enables easy ad hoc reconstructions of PES that are tailored to a particular problem at hand and we anticipate that this will enable new insights when studying complex physical interactions in situations were the true \emph{ab initio} calculations are prohibitively expensive (e.g. CCSD(T) MD).

Due to its global formulation, sGDML is able to account for the full scope of atomic interactions within the studied system, alas at the cost of transferability. A model trained on conformers of one molecule can not be used to infer about energies and forces of another. However, one particular challenge that seems to be achievable without sacrificing globality, is to train unified models for larger molecular families, e.g. paracetamol and methyl-substituted paracetamol, etc.

\section{Software Availability}

Our code and documentation is available at \texttt{http://sgdml.org}.

\section{Acknowledgements}
S.C., A.T., and K.-R.M. thank the Deutsche Forschungsgemeinschaft (projects MU 987/20-1 and EXC 2046/1 [ID: 390685689]) for funding this
work. A.T. is funded by the European Research Council with ERC-CoG grant BeStMo. This work was supported by the German Ministry for Education and Research as Berlin Big Data Centre (01IS14013A) and Berlin Center for Machine Learning (01IS18037I). This work was also supported by the Information \& Communications Technology Planning \& Evaluation (IITP) grant funded by the Korea
government (No. 2017-0-00451). This publication only reflects the authors views. Funding agencies are
not liable for any use that may be made of the information contained herein. Part of this research was performed while
the authors were visiting the Institute for Pure and Applied Mathematics, which is
supported by the NSF.

\bibliography{references/clean_ref.bbl}

\end{document}